\documentclass[12pt]{article}

\def\be{\begin{equation}}
\def\ee{\end{equation}}
\def\ApJ{\sl ApJ}
\def\MNRAS{\sl MNRAS}
\def\AA{\sl A\&A}
\def\AJ{\sl AJ}
\def\ARAA{\sl ARA\&A}

\title{\bf{ Constraints on Cosmic Equation of State using Gravitational
Lensing Statistics with evolving galaxies}} 
\author{Abha Dev\thanks{E--mail : abha@ducos.ernet.in},
Deepak Jain\thanks{Also at Deen Dayal Upadhyaya college, University of
Delhi, Delhi-110 015 },
N. Panchapakesan,\\
S. Mahajan and
V. B. Bhatia\\
	{\em Department of Physics and Astrophysics} \\
        {\em University of Delhi, Delhi-110 007, India} 
	}

\usepackage{overcite}
\begin {document}
\maketitle

\begin{center}
\Large{\bf Abstract}
\end{center}
\large
\baselineskip=20pt
  %~~~~~~~~~~~~~~~~~~~~~~~~~~~~~~~~~~~~~~~~~~~~~~~~~~~~~~~
%                ABSTRACT
%~~~~~~~~~~~~~~~~~~~~~~~~~~~~~~~~~~~~~~~~~~~~~~~~~~~~~~~

In this paper, observational constraints on the cosmic 
equation of state of dark energy ($p = w \rho$) have been investigated
using gravitational lensing
statistics. A likelihood analysis of the lens surveys has been carried
out to constrain the cosmological parameters $\Omega_{m}$ and $w$.  
Constraints on $\Omega_{m}$ and $w$ are obtained in 
three different models of galaxy evolution: no evolution model
(comoving number density of galaxies remain constant), Volmerange and
Guiderdoni model and fast merging model. The last two models consider
the number evolution of galaxies in addition to the luminosity
evolution. The likelihood analysis shows that for the no-evolution
case  $w \leq -0.04$ and   
$\Omega_{m}\leq 0.90$ at $1\sigma$ ($68\%$ confidence level).
Similarly for the Volmerange $\&$ Guiderdoni Model the constraints are 
$w \leq -0.04$ and $\Omega_{m} \leq 0.91$ 
at $1 \sigma$. In fast merging model the constraint become weaker
and it allows almost
the entire range of parameters.  
For the case of constant $\Lambda$ ($w =-1$), all the models permit 
$\Omega_{m} = 0.3$ with $68\%$ CL which is consistent with the value
of $\Omega_{m} $ inferred from various other cosmological observations.

      \vfill
      \eject

%\Huge

\baselineskip = 20pt
\begin {section} {Introduction}
   %~~~~~~~~~~~~~~~~~~~~~~~~~~~~~~~~~~~~~~~~~~~~~~~~~~~~~~~~~~
%                  INTRODUCTION
%~~~~~~~~~~~~~~~~~~~~~~~~~~~~~~~~~~~~~~~~~~~~~~~~~~~~~~~~~

Recent observations are in concordance with flat cosmological models in
which the universe is in an accelerating phase. Analysis of the magnitude-redshift
data of high-redshift type Ia supernovae (SNe Ia) suggests that
the ratio of the matter energy to the critical energy, $\Omega_{m}$, is $\sim
0.3$ \cite{Perl,Bperl,Riess}. SNe Ia calibrated ``standard'' candles
appear fainter  
than would be expected if the expansion was slowing due to
gravity. Recent studies of the Cosmic Microwave Background (CMB) data
also favour a nearly flat universe: the first acoustic peak in the angular
power spectrum of the CMB is located at $l\sim (197\pm 6)$
\cite{Bernardis}. Theoretical modeling of structure formation based on
cold dark matter models (CDM) with $\Omega_{m} = 1$  fails to
reconcile with the observations at the quantitative level. 
By contrast, a spatially
flat $\Lambda$ CDM universe (non-zero cosmological constant) with 
$\Omega_{m} \simeq 0.3$ can explain observations of galaxy clustering on
large scales, increases the age of the universe (which helps to
accommodate the age of globular clusters), and makes the total
energy density equal to critical density as generically predicted by
inflationary cosmology \cite{OS,Varun}.

Neither observational evidence, nor inflationary considerations tell
us that the cosmological term (dominant, negative pressure term) is a
constant. There are various  phenomenological models of
dynamical-$\Lambda$ (dark       
energy component) present in the literature. These are:
% can be classified into
%three groups \cite{Varun}:

\begin{enumerate}
\item $\Lambda$ varies with cosmic time or with the scale factor of
Friedmann-Robertson-Walker (FRW) metric  
\cite{Bloom,Silveira,vs,Waga}. 

\item An X-matter cosmology where this unknown form of energy is
characterised by an  equation
of state $p_{x} = w\rho_{x}$ \cite{msTurner,ht,Chiba,TChiba}.   

\item Rolling scalar field models (quintessence):
The $\Lambda$ term is considered to be a new, physical, classical field with
some phenomenological Lagrangian. In this class the most popular
models are those of an evolving scalar field that couples minimally to
gravity \cite{Peebles,Ratra,Caldwell}.
\end{enumerate}
 
In the present work, we focus our attention on the dark energy 
component characterised by the equation of state, namely the ratio $w =
p/\rho$. In particular, we work with an equation of state with $-1\leq w \leq 0$
because this range fits current cosmological observations best. The 
cosmological constant is a special case of this equation of state
corresponding to $w = -1$. 
Constraints from large scale structure (LSS) and CMB complemented by
the SNe Ia data require $0.6 \leq \Omega_{x} \leq 0.7$ and $w < -0.6$
for a flat universe \cite{Bperl,gef} and $w <-0.4$ for universes with
arbitrary spatial curvature \cite{gef}. Alcaniz and Lima (1999)
\cite{Lima} have derived the limits on the cosmic equation of state from age
measurements of old high redshift galaxies (OHRG). They show that 
if, as indicated from dynamical measurements,
the density parameter $\Omega_{m} \sim 0.3$, then $w \leq -0.27$. 
By combining ``cosmic concordance'' with
maximum likelihood estimator, Wang \emph{et al.} (2000)\cite{wa} 
find that the best-fit model lies in
the range $\Omega_{m}=  0.33\pm 0.05$ with an effective equation of state
$w \sim -0.65\pm 0.07$. 

In this work we use the statistics of gravitational lensing of quasars
to set 
quantitative limits on the present density of dark energy component
and matter.
 
The main aim of this paper is to constrain the cosmic
equation of state using
gravitational lensing statistics as a tool, taking into account the {\it
evolution of lenses (galaxies)}. Many observations suggest 
that the galaxies we see
today could have evolved from the merging of smaller subsystems. Hence,
the inclusion of the fact that the number density of lensing galaxies
changes with time is very important in the lensing statistics\cite{rix,ma,dj}. 
We consider two different evolutionary models of galaxies : the
fast merger model of Broadhurst, Ellis \& Glazebrook (1992) \cite{bro} 
and the Volmerange \& Guiderdoni model (here after VG model)
\cite{rocca}. Both the  
models consider the number evolution of galaxies in 
addition to the pure luminosity evolution.

The paper is organized as follows. In Section 2, we review the 
``dark energy component'' model. In Section 3, we discuss
the two evolutionary models we are considering. 
In Section 4, we write down
the lensing probabilities for the three models: non-evolutionary model
(where the comoving number density of lensing galaxies remains
constant), fast-merging and VG model. We also
discuss the likelihood analysis of the lens surveys. Section 5 is 
devoted to a discussion of the results.

   \end {section}

\begin {section} {Field Equations and Distance Formula}
   %~~~~~~~~~~~~~~~~~~~~~~~~~~~~~~~~~~~~~~~~~~~~~~~~~~~~~~~~~~~~
%             DARK ENERGY MODEL
%~~~~~~~~~~~~~~~~~~~~~~~~~~~~~~~~~~~~~~~~~~~~~~~~~~~~~~~~~~~~

For spatially flat, homogeneous, and isotropic cosmologies with 
non-relativistic matter and a separately conserved ``dark energy component''
with equation of state $p = w\rho$, the Einstein field equations are:

\be
(\frac{\dot{a}}{a})^{2} = H_{o}^{2}\,[\Omega_{m}(\frac{a_{o}}{a})^{3} + \Omega_{x}(\frac{a_{o}}{a})^{3(1+w)}]
\ee
\noindent and
\be
\frac{\ddot{a}}{a} = - \frac{H_{o}^{2}}{2} \,[\Omega_{m}(\frac{a_{o}}{a})^{3} + (3w+1)\Omega_{x}(\frac{a_{o}}{a})^{3(1+w)}],
\ee

\noindent where the overdot denotes derivatives with respect to time, $H$
is the Hubble parameter and $a$ is the scale factor.
Subscript $o$ refers to present values. $\Omega_{m}$ and $\Omega_{x}$
are present day matter and dark energy density parameters. 
As we are mostly interested in effects that occured at redshift 
$z < 5$, we neglect the energy density of radiation. The   
condition for the flat universe becomes $\Omega_{x} = 1 -
\Omega_{m}$. For more details see for example Zhu (2000)\cite{zh}

The proper distance between a source at redshift z and the observer at
$z = 0$ is given by

\be
d(0,z) = a_{o} \chi(z) \nonumber 
       = c\,H_{o}^{-1}\,\int_{0}^{z} \frac{dy}{\sqrt{\Omega_{m}(1+y)^{3} + 
(1-\Omega_{m})(1+y)^{3(1+w)}}},
\ee

\noindent where $\chi(z)$ is the comoving distance to the source.

The angular diameter distance for our models, between redshift $z_{L}$ and
$z_{S}$ reads:

\be
d_{A}(z_{L},z_{S}) = \frac{c\,H_{o}^{-1}}{1+z_{S}}\,\int_{z_{L}}^{z_{S}} \frac{dy}{\sqrt{\Omega_{m}(1+y)^{3} + (1-\Omega_{m})(1+y)^{3(1+w)}}}.
\ee

   \end {section}

\begin {section} {Evolution Of Galaxies}
   %~~~~~~~~~~~~~~~~~~~~~~~~~~~~~~~~~~~~~~~~~~~~~~~~~~~~~~~~~~~
%                EVOLUTION OF GALAXIES
%~~~~~~~~~~~~~~~~~~~~~~~~~~~~~~~~~~~~~~~~~~~~~~~~~~~~~~~~~~~
      
Gravitational lens statistics and galaxy evolution are linked with each
other because the galaxies which are evolving through different
mechanisms  are basically acting as lenses. 
Therefore gravitational lens
statistics is certainly going to be affected if the lenses 
evolve. The theory of the formation and evolution
of galaxies 
is one of the major unsolved problems of astrophysics. According to
one view,
galaxies evolved through a complex series of interactions and have
settled in the present day form \cite{too,schw}. Others believe
that galaxies
were created in a well defined event at a very early time in the
history of the universe\cite{egg,pa}. It remains
unclear which process dominates the formation of elliptical
galaxies.
Among the many theories of galaxy formation, the idea that galaxies may
form by
the accumulation of smaller star-forming subsystems has recently
received much attention. 

\noindent Observational evidence also supports this
``bottom - up'' scheme.  
Deep Hubble Space Telescope (HST) images\cite{dri} and the
size-redshift
relation of luminous early-type galaxies (E/S0) and mid-type spiral
galaxies (Sabc)
indicate that these objects were assembled largely before $z = 1$ 
and have been evolving passively for $ z < 1$. Moreover, HST 
and
ground-based telescopes show that the galaxy-merger rate was higher
in the past and it roughly increases with
redshift\cite{bur0,Faber,carl}. These arguments
suggest that the galaxies we see today could have been assembled
from
the merging of smaller systems sometime during $z > 1$.

%known observationally.
Recent observations \cite{z} show that
there is a strong deficit of galaxies with extreme red
colors (seen in
different separated fields and at different flux limits) than
predicted by
models in which elliptical galaxies completed their star formation
by
$z \sim 5$. Therefore, elliptical galaxies must have had
significant
star formation at $ z < 5$ through merging and associated
star bursts. The formation of elliptical galaxies in this way is
also
consistent with the predictions of hierarchical clustering models
of
galaxy formation.

The other piece of evidence which supports the merging hypothesis comes
from the excess of Faint Blue
Galaxies [FBG] which have been found in many deep imaging studies
\cite{gla,eil,lil,reil}. Comparison with a model which assumes no
luminosity evolution in
the galaxy population shows that at B = 24, the actual galaxy
count
exceeds the model predictions by a factor of 5. Merging of galaxies
can
also solve the surprisingly steep increase in the number density of
galaxies\cite{rocca,gui,bro}. It is also found that the ``FBG
problem'' cannot be resolved in the conventional scenario of formation
and evolution of ellipticals. In this scenario, elliptical galaxies
are assumed to have formed in an instantaneous
burst of star formation at high redshifts with no subsequent star
formation events.

Although galaxy mergers are no longer a matter of dispute,
there is no agreement on the current and past rate of galaxy
mergers. Several authors have attempted to find the
dependence of merger rate on the redshift \cite{ze,yee}. 
There are many
challenges for the  mergers theory  which may be met in the
near future with more observations.
Recently there have been several semi-analytic models,
motivated by cosmological theory, which may eventually provide a
greater understanding of galaxy formation and evolution
\cite{cole,bau0,bau1,bau2}. The
traditional
galaxy number-count models, however, are still powerful tools in
exploring the formation and evolution of galaxies and can be
treated
as complements to the semi-analytic techniques.

To explain the galaxy
number-counts various number-luminosity evolution models have been
constructed. Along with the no-evolution model, we consider two
different models of galaxy evolution
which try to explain some of the observational facts listed
above. These models are: 

\begin{enumerate}
\item VG model \cite{rocca}. 
\item Fast merging model \cite{bro}.
\end{enumerate} 

The general philosophy behind these merger models is to assume
that the current giant galaxies result from the merging of a number
of smaller ``building blocks''. It is found that under some general
assumptions the theoretical predictions of merging models nicely fit
the observed galaxy counts. The assumptions underlying these models are: 

\begin{enumerate}
\item Stability of the Schechter form for the luminosity function.
\item Conservation of total comoving mass density. 
\end{enumerate}

The evolving luminosity function at any
redshift z is given as

\be
\Phi(L,z) \,dL=
\phi_\ast(z)\,\, \left({L \over L_\ast(z)}\right )^\alpha\,\, exp\left (-
{L\over L_\ast(z)}\right )\,\, {dL\over L_\ast(z)}
\label{lf1}
\ee

\noindent $L_{\ast}(z)$ being the characteristic luminosity at the
knee, $\phi_{\ast}(z)$ a characteristic density and $\alpha$ is the
index of faint end slope.
\vskip 0.5cm
{\bf{\underline{No-Evolution Model}}}
\vskip 0.4cm
\noindent This model assumes that the comoving number density of galaxies
is constant and the mass of galaxies does not change with  cosmic
time. Therefore
\be
\phi_{\ast}(z) =  \phi_{\ast}(0) = \mathrm{constant}
\ee

\noindent The characteristic luminosity at any redshift remains constant
and hence the mass of galaxy at any redshift is constant.

\begin{equation}
L_{\ast}(z) = L_{\ast}(0) = \mathrm{constant}
\end{equation}

\noindent The  ``$0$'' refers to  present-day values.

\noindent The Schechter form is the most commonly used luminosity
function for the lens galaxies. Therefore eq.(\ref{lf1}) becomes 
%of the  Schechter
%form \cite{sch}  
\begin{equation}
\Phi(L,z) \,dL \,= \,\Phi(L, z = 0)dL \,=\,
\phi_\ast\, \left({L \over L_\ast}\right )^\alpha \,\,exp\left (-
{L\over  L_\ast}\right ) \,{dL\over L_\ast}
\label{seht}
\end{equation}

\noindent where $\phi_{*}$, $\alpha$ and $L_{*}$ are the
normalization factor, the index of faint-end slope and the
characteristic luminosity at the present epoch respectively. These values
are fixed in order to fit the current luminosities and densities of
galaxies. This is known as the  Schechter
form of the luminosity function\cite{sch}.

\vskip 0.5cm
\noindent{\bf{\underline{VG  Model}}}
\vskip 0.5cm
In 1990, Volmerange and Guiderdoni \cite{rocca}, proposed a
unifying
model to
explain faint galaxy counts as well as observational properties of
distant radio galaxies. This model of galaxy evolution 
is based on number evolution in addition to pure luminosity evolution.
According to this model the present day
galaxies result from the merging of a large number of building
blocks and the comoving number of these building blocks evolves as $ ( 1 +
z)^{1.5}$.

It is argued that the present luminosity function is the
well known Schechter Luminosity Function \cite{sch} given in
eq.(\ref{seht}) above. Then at
high z, the galaxies follow a new luminosity
function where $\phi_\ast(z)$ and $L_\ast(z)$  vary as:
\be
\phi_\ast(z)\, =\,\phi_\ast(0)\, (1 + z)^\eta
\ee
\noindent and
\be
L_\ast(z)\, =\, L_\ast(0)\,(1 + z)^{-\eta}
\ee
eq.(\ref{lf1}) now becomes
\begin{equation}
\Phi(L,z)\,dL = (1 +
z)^{2\eta}\,\phi_{\ast}\,\,\left[{L\over L_{\ast}}(1 +
z)^{\eta}\right]^{\alpha}\,\, exp\left[ {-L\over{L_{\ast}}}(1 +
z)^{\eta}\right]\,{dL\over{ L_{\ast}}}
\end{equation}
\noindent or
\begin{equation}
\Phi(L,z)\,dL\, =\, ( 1 + z)^{2\eta}\,\, \Phi(L(1 + z)^\eta, z = 0)\, dL.
\end{equation}

\noindent It is seen that the value $\eta = 1.5$ gives a fair fit to
the data on high redshift galaxies. The functional form has the
following properties:
\begin{enumerate}
\item [(i)] Self-similarity as suggested by the  Press-Schechter
formalism subject to the constraint that the total comoving mass of
associated material is conserved  \cite{press1}.

\item [(ii)] The comoving number density evolves as
$\phi_\ast(z)\,=\,\phi_\ast(0)\,(1 + z)^\eta$ and the
characteristic
luminosity of the galaxy luminosity function varies
as $L_\ast(z)\, =\, L_\ast(0)\,(1 + z)^{-\eta}$. Thus the galaxies are
more numerous and less massive for increasing z.
\end{enumerate}

\vskip 0.5cm

\noindent {\bf{\underline{Fast Merging Model}}}
\vskip 0.5cm
The fast merger model of
Broadhurst, Ellis \& Glazebrook (1992) [BEG] was originally motivated
by the faint galaxy population counts \cite{bro}. This model also
assumes the comoving number density of the lenses to be a function of
the look back time $\delta t$ or redshift $z$ and hence:

\begin{equation}
\phi_\ast(\delta t) = f(\delta t)\,\, \phi_\ast(0)
\end{equation}

\noindent and

\begin{equation}
L_\ast(\delta t) = [f(\delta t)]^{-1}\, L_\ast(0).
\end{equation}

\noindent Since luminosity is related to the velocity dispersion 
of the dark halo of the lensing galaxy through the
Faber-Jackson relation $( L \propto v^{\gamma})$ \cite{Faber}, 
this form implies
that if we had $n$ galaxies at  time 
$\delta t$ each with velocity dispersion $v$, they would by today
have
merged into one galaxy with a velocity dispersion $[f(\delta t)]^{1
\over \gamma}v$. The strength and the time dependence of merging is
described by the function $f(\delta t)$:

\begin{equation}
f(\delta t) = exp\,(Q\, H_{0}\,\delta t)
\end{equation}

\noindent where $H_{0}$ is the Hubble constant at the present epoch and Q
represents the merging rate. As suggested by BEG \cite{bro}, we take
Q = 4. The 
look back time $\delta t$ is related to the redshift  $z$ through

\begin{equation}
H_{0}\delta t = \int^{z}_{0}\frac{(1+y)^{-1}\,dy}{\sqrt{\Omega_{m}(1 + y)^{3}+(1-\Omega_{m})(1+y)^{3(1+w)}}}.
\end{equation}

\noindent Now we can rewrite eq.(\ref{lf1}) in terms of $\delta t$
instead of $z$.
\begin{equation}
\Phi(L,\delta t)\,dL = f(\delta t)^{2}
\,\phi_{\ast}\,\,\left[{L\over L_{\ast}}f(\delta t)
\right]^{\alpha}\,exp\left[ {-L\over{L_{\ast}}}f(\delta t)
\right]\,{dL\over{ L_{\ast}}}
\end{equation}
\noindent or
\begin{equation}
\Phi(L,\delta t)\,dL = f(\delta t)^{2}\,\,\,[ \,\,\Phi(\,\,L\,f(\delta
t),\,\, z = 0 
\,)\, \,] \,dL   
\label{lf2}
\end{equation}

   \end {section}

\begin{section} {Likelihood Analysis of Lens Surveys}
   %~~~~~~~~~~~~~~~~~~~~~~~~~~~~~~~~~~~~~~~~~~~~~~~~~~~~~~~~~~~~
%      Likelihood Analysis and Gravitational Lensing
%~~~~~~~~~~~~~~~~~~~~~~~~~~~~~~~~~~~~~~~~~~~~~~~~~~~~~~~~~~~~

\subsection{\emph{Basic Equations of Gravitational Lensing Statistics}}
\vskip 0.5cm
For simplicity we use the Singular Isothermal Model (SIS) 
for the lens mass distribution.
The cross-section for lensing events for the SIS model is given by \cite{TOG} 

\be
\sigma = 16\,\pi^{3}\,({v \over c})^{4}(\frac{D_{OL}D_{LS}}{D_{OS}})^{2}, 
\label{sigma}
\ee

\noindent where $v$ is the velocity dispersion of the dark 
halo of the lensing galaxy, $D_{OL}$ is the angular diameter 
distance to the lens, $D_{OS}$ is the
angular diameter distance to the source and $D_{LS}$ is the angular diameter
distance between the lens and the source.
The mean image separation for the lens at $z_{L}$ takes a simple form 

\begin{equation}
\label{theta}
\Delta\theta = 8\,\pi\,(\frac{v}{c})^{2}\,\frac{D_{LS}}{D_{OS}}.
\end{equation}

The differential probability $d\tau$ of a beam having a lensing 
event in traversing $dz_{L}$ is

\be
d\tau = n_{L}(z)\,\sigma \,\frac{cdt}{dz_{L}} dz_{L}, 
\label{dtau1}
\ee
\noindent where $n_{L}(z)$ is comoving number density of the lenses
 and the quantity $cdt/dz_{L}$ is calculated to be

\be
\frac{cdt}{dz_{L}} = \frac{a_{o}}{(1 + z_{L})} \,\frac{1}{\sqrt{\Omega_{m}
(1 + z_{L})^{3} + (1 - \Omega_{m})(1 + z_{L})^{3(1+w)}}},
\ee

\noindent where $a_{o}$ is scale factor at the present epoch.
Substituting for $\sigma$ from eq. (\ref{sigma}), we get
 
\be 
d\tau = n_{L}(z)\left[ {16 \pi^{3}\over {c H_{0}^{3}}}
v^{4}({D_{OL}D_{LS}\over a_{o} 
D_{OS}})^{2}{1\over a_{o}}\right ]{cdt\over dz_{L}}dz_{L}. 
\label{ro1}
\ee

%%~~~~~~~~~~~~~~~~~~~~~~~~~~~~~~~~~~~~~~~~~~~~~~~~~~~~~~~~%%%
\vskip 0.5cm
\noindent{\bf The No-Evolution Model}

\noindent The luminosity function is given by eq. (\ref{seht}) and therefore we have
\be
< n_{L}(z)\, v^{4}> =
(1 + z_L)^{3} \, v^{4}_{*}\int_{0}^{\infty}\Phi(L)dL\, \left({v\over
v_{\ast}}\right)^4.
\label{ro3}
\ee

\noindent We assume that $v$ is linked with $L$ through the Faber-Jackson relation
for elliptical galaxies
\begin{equation}\left({L\over
{L_{\ast}}} \right) = \left({v\over v_{\ast}}\right)^{\gamma}.
\label{ro4}
\end{equation}

\noindent Hence eq.(\ref{ro3}) becomes
\begin{equation}
< n_{L}(z)\, v^{4} > =
(1 + z_L)^{3} \, v_{*}^{4}\int_{0}^{\infty}\Phi(L)dL\left({L\over
{L_{\ast}}} \right)^{4\over \gamma}.
\label{ro5}
\end{equation}

\noindent The optical depth can be written as
\be
d\tau = F^{*}(1+z_L)^{3}\left({{D_{OL} D_{LS}}\over{ a_{o} D_{OS}}}\right)^{2}
{1\over a_{o}}\,{c dt\over dz_{L}} dz_{L}
\label{tau},
\ee 

\noindent with

\be
F^* = {16\pi^{3}\over{c
H_{0}^{3}}}\phi_\ast v_\ast^{4}\Gamma\left(\alpha + {4\over\gamma} +1\right).
\ee

\noindent $F^{*}$ is a dimensionless quantity which governs the 
probability of a beam encountering a lensing object. It is a measure 
of the effectiveness of matter in producing multiple images \cite{TOG}. 
For our calculation we use Schechter and lens parameters for E/SO
galaxies as suggested by J. Loveday et al. (hereafter $LPEM$
parameters) \cite{loveday}:
$\phi_{\ast} = 3.2\pm 0.17 h^{3}\,10^{-3} Mpc^{-3}$, 
$\alpha = 0.2$, $\gamma = 4$, $v_{\ast} = 205.3\,km/s$ and $F^{*} = 0.010$.
\vskip .2cm

\noindent The differential optical depth of lensing in traversing
$dz_{L}$ with angular separation between $\phi$ and $\phi + d\phi$, in
the presence of evolution of galaxies, is
given by

\begin{eqnarray}
\frac{d^{2}\tau}{dz_{L}d\phi}d\phi dz_{L} 
&=& F^{*}\,(1 + z_{L})^{3}\,\left({{D_{OL} D_{LS}}\over{ a_{o} D_{OS}}}\right)^{2}\,\frac{1}{a_{o}}\, \frac{cdt}{dz_{L}}\,\frac{\gamma/2}{\Gamma(\alpha+1+\frac{4}{\gamma})} \nonumber \\
&&\left(\frac{D_{OS}}{D_{LS}}\phi\right)^{\frac{\gamma}{2}(\alpha+1+\frac{4}{\gamma})}\,exp[-\left(\frac{D_{OS}}{D_{LS}}\phi\right)^{\frac{\gamma}{2}}] \frac{d\phi}{\phi} dz_{L}
\label{diff}
\end{eqnarray}

\noindent for ellipticals (lenticulars), where $\phi = \Delta\theta/8\pi (v_{\ast}/c)^{2}$ with $v_{\ast}$ the velocity dispersion corresponding to the characteristic
luminosity $L_{\ast}$ in (\ref{ro4}).

%%~~~~~~~~~~~~~~~~~~~~~~~~~~~~~~~~~~~~~~~~~~~~~~~~~~~~~~~~%%% 

\vskip 0.5cm
\noindent{\bf The Evolutionary Models}

\noindent In the case where the comoving number density of the lenses
changes with time, the equations (\ref{ro5}), (\ref{tau}) and (\ref{diff}) read as     \cite{dj} 
\be
< n_{L}(z)\, v^{4} > =
(1 + z_L)^{3} \, v_{*}^{4}\int_{0}^{\infty}\Phi(L,z)dL\left({L\over
{L_{\ast}}} \right)^{4\over \gamma},
\label{ero5}
\ee

\be
d\tau = F^{*}(1+z_L)^{3}\left({{D_{OL} D_{LS}}\over{ a_{o} D_{OS}}}\right)^{2}
\,\Psi(\delta t)^{(1-\frac{4}{\gamma})}\;{1\over a_{o}}{c dt\over dz_{L}} dz_{L}
\label{etau},
\ee

\noindent and

\begin{eqnarray}
\frac{d^{2}\tau}{dz_{L}d\phi}d\phi dz_{L} 
&=& F^{*}\,(1 +z_{L})^{3}\,\left(\frac{D_{OL}D_{LS}}{a_{o}D_{OS}}\right)^{2}\,\frac{1}{a_{o}}\,\frac{cdt}{dz_{L}} \nonumber \\ 
&&\times \Psi(\delta
t)^{2+\alpha}\,\frac{\gamma/2}{\Gamma(\alpha+1+\frac{4}{\gamma})}\,\left(\frac{D_{OS}}{D_{LS}}\phi\right)^{\frac{\gamma}{2}(\alpha+1+\frac{4}{\gamma})}\nonumber \\
&&exp\left [-\left(\frac{D_{OS}}{D_{LS}}\phi\right)^{\frac{\gamma}{2}}\,\Psi(\delta t)\right]\,\frac{d\phi}{\phi}\,dz_{L},
\label{ediff}
\end{eqnarray}

\noindent where $\Psi(\delta t) = exp(Q\;H_0 \delta t)$ for the fast merging
model and $\Psi(\delta t) = (1+z)^{1.5}$ for the VG model for
evolution of lensing galaxies.
We notice that when  $\gamma = 4$ the differential probability is the
same for both 
evolutionary model and no evolutionary model.

Lensing increases the apparent brightness of a quasar causing
over-representation of multiply imaged quasars in a
flux-limited sample. This effect is called the magnification bias. The
bias factor for a quasar at redshift $z$ with apparent magnitude $m$ is 
given by \cite{Turner,Fuku,FFKT,1CSK,2CSK}

\be
{\bf B}(m,z) = M_{0}^{2}\, \emph{B}(m,z,M_{0},M_{2}),
\label{bias}
\ee 

\noindent where

\be
\emph{B}(m,z,M_{1},M_{2}) =
2\,\left(\frac{dN_{Q}}{dm}\right)^{-1}\int_{M_{1}}^{M_{2}}\frac{dM}{M^{3}}\,\frac{dN_{Q}}{dm}(m+2.5\log(M),z).
\label{bias1}
\ee

\noindent In the above equation
$({dN_{Q}(m,z)}/{dm})$ is the measure of number of quasars with magnitudes 
in the interval $(m,m+dm)$ at redshift $z$.

We can allow the upper magnification cut off $M_{2}$ to be infinite,
though in practice we set it to be $M_{2} = 10^{4}$. $M_{0}$ is the
minimum magnification of a multiply imaged source and for the SIS model 
$M_{0} =2$.

We use Kochanek's ``best model'' \cite{2CSK} for the quasar luminosity
function:

\be
\frac{dN_{Q}}{dm}(m,z) \propto
(10^{-a(m-\overline{m})}+10^{-b(m-\overline{m})})^{-1}
\label{lum}.~~~~~~~~~~~~~~~~~~~~
%                        RESULTS
\ee

\noindent where the bright-end slope $a$ and faint-end slope $b$ are
constants, and the break magnitude $\overline{m}$ evolves with redshift:   

\be
\overline{m} = \left\{ \begin{array}{ll}
                   m_{o}+(z-1)    & \mbox{for $z < 1$}, \\
                   m_{o}          & \mbox{for $1 < z \leq 3$}, \\
                   m_{o}-0.7(z-3) & \mbox{for $z > 3$}.
                   \end{array}
               \right. \
\ee
 
\noindent Fitting this model to the quasar luminosity function data in
Hartwich $\&$ Schade \cite{Qdata} for $z > 1$, Kochanek finds that
``the best model'' has $a = 1.07 \pm 0.07$, $b = 0.27 \pm 0.07$
 and $m_{o} = 18.92 \pm 0.16$ at B magnitude.    

The magnitude corrected probability, $p_{i}$, for the quasar i with
apparent magnitude $m_{i}$ and redshift $z_{i}$ to get lensed is:

\be
p_{i} = \tau(z_{i}){\bf B}(m_{i},z_{i}).
\label{prob1}
\ee

%We use $ x = 1.07$, $y = 0.27$ and $m_o = 18.92$.
% We considered a total of 862
% $ (z > 1)$ highly  luminous optical quasars plus five lenses.

 Selection effects are caused by limitations on dynamic range,
 limitations on resolution and presence of confusing sources such as
 stars. Therefore we must include a selection function to correct
 the
 probabilities. In the SIS model the selection function is modeled by
 the
 maximum magnitude difference $\Delta m(\theta)$  that can be detected for
 two images
 separated by $\Delta\theta$. This is equivalent
 to
 a limit on the flux ratio $( f > 1)$ between two images $ f =
 10^{0.4\,\Delta m(\theta)}$. The total magnification of images
 becomes
 $M_{f} = M_{0}(f+1)/(f-1)$. The survey can only detect lenses
 with
 magnifications larger than $M_{f}$. This sets the lower limit on the
 magnification. Therefore the $M_{1}$ in the bias function \ref{bias1} gets replaced
 by $M_{f}(\theta)$. 
 The corrected lensing probability and image separation distribution function
 for a single source at redshift $z_{S}$ are given as \cite{2CSK}

\be
p^{'}_{i}(m,z) = p_{i}\int \frac{ d(\Delta\theta)\,
p_{c}(\Delta\theta)\emph{B}(m,z,M_{f}(\Delta\theta),M_{2})}{\emph{B}(m,z,M_{0},M_{2})},
\label{prob2}
\ee

\noindent and

\be
p^{'}_{ci} =
p_{ci}(\Delta\theta)\,\frac{p_{i}}{p_{i}^{'}}\,\frac{\emph{B}(m,z,M_{f}(\Delta\theta),M_{2})}
{\emph{B}(m,z,M_{0},M_{2})},
\label{confi}
\ee

\noindent where

\be
p_{c}(\Delta\theta) =
\frac{1}{\tau(z_{S})}\,\int_{0}^{z_{S}}\frac{d^{2}\tau}{dz_{L}d(\Delta\theta)}
\,dz_{L},
\label{pcphi}
\ee

%\be
%M_{f}(\Delta\theta) = \begin {array}{ll}
%                   M_{0}\,(\frac{f+1}{f-1}) & \mbox{for $f > 1$,}
%                 \end {array}
%\ee 

%\noindent and

%\be
%f = f(\Delta\theta) = 10^{0.4\Delta m(\Delta\theta)}.
%\ee

\noindent Equation (\ref{confi}) defines the configuration
probability. It is the probability that the lensed quasar i is lensed
with the observed image separation.  
%$p_{ci}(\Delta\theta)$ in eq. (\ref{confi}) 
%is $p_{c}(\Delta\theta)$ evaluated for the observed image separation.

To get selection function corrected probabilities, we divide our
sample into two parts - the ground based surveys and the HST survey. We
use the selection functions as suggested by Kochanek \cite{1CSK}.

In our present calculations we do not consider the extinction effects
due to the presence of dust in the lensing galaxies.

The image separation distribution, $dN/d(\Delta\theta)$, is given by
\cite{c,DA} 

\be
\frac{dN}{d(\Delta\theta)} = \sum_{i =1}^{N} \,{\mathcal{P}}_{i}(\Delta\theta)
\ee

\noindent where N is the total number of optical quasars considered and 

\be
{\cal {P}}_{i}(\Delta\theta) 
= M_{0}^{2}\,\emph{B}(m_{i},z_{i},M_{f}(\theta),M_{2})\int_{0}^{z_{S}} \frac{d^{2}\tau}{dz_{L}d(\Delta\theta)}\;dz_{L}.
\ee

The sum of the lensing probabilities $p_{i}^{'}$ for the optical QSOs gives the
expected number of lensed quasars, $N_{lens} = \sum\, p_{i}^{'}$. The
summation is over the given quasar sample.
 
\vskip1.0cm
\subsection{\emph {Testing the models against observations}}
\vskip 0.5cm

We perform maximum-likelihood analysis to determine the confidence
level of $w$ and $\Omega_{m}$ for the case where comoving number density of lenses
is constant and for the two evolutionary models, namely, the VG model
and the fast-merging model.

The likelihood function is

\be
{\cal{L}} = \prod_{i=1}^{N_{U}}(1-p^{'}_{i})\,\prod_{k=1}^{N_{L}}
p_{k}^{'}\,p_{ck}^{'}.
\label{LLF}
\ee

\noindent Here $N_{L}$ is the number of multiple-imaged lensed
quasars, $N_{U}$ is the number of unlensed quasars, $p_{k}^{'}$, 
the probability of quasar k to get lensed, is given by eq. (\ref{prob2})
and $p_{ck}^{i}$, the configuration probability, is given by eq. (\ref{confi}). 
We considered a total of 867 ($z > 1$) high luminosity optical quasars 
which include 5 lenses. These are taken from optical lens surveys such as the
HST Snapshot survey \cite{HST}, the Crampton survey \cite{Crampton}, 
the Yee survey \cite{Yee},Surdej survey\cite{Surdej}, the NOT Survey
\cite{Jaunsen} and the FKS survey \cite{FKS} .
 The lens surveys and quasar catalogs usually use V
magnitudes, so we transform $m_{V}$ to a B-band magnitude using $B-V =
0.2$  as suggested by Bachall et al. (1992)\cite{Bahcall}.

   \end {section}

\begin{section}{Results And Discussion}
% ~~~~~~~~~~~~~~~~~~~~~~~~~~~~~~~~~~~~~~~~~~~~~~~~~~~~~~
%                        RESULTS
%~~~~~~~~~~~~~~~~~~~~~~~~~~~~~~~~~~~~~~~~~~~~~~~~~~~~~~

 The likelihood function as defined in eq. (\ref{LLF}) is a function
 of the parameters $\Omega_{m}$ and $w$. We allow the parameters
 $\Omega_{m}$ and $w$ to vary in the range $0 \leq \Omega_{m} \leq
 1$ and
 $-1\leq w \leq 0$ and find the maximum of the likelihood
 ${\mathcal{L}}_{max}$. The
 logarithm of the ratio of the likelihood to its maximum
 $-2\ln({\mathcal{L}}/{\mathcal{L}}_{max})$ is asymptotically
 distributed like a $\chi^{2}$
 distribution with the degrees of freedom equal to the parameters
 involved. In the figures we present likelihood function as a
 function
 of two parameters and therefore the $68\%$($1 \sigma$) and $95.4\%$
 ($2
 \sigma$) confidence levels are defined by the contours on which
 $\cal{L}$
 is $31.6\%$ and $4.5\%$ of ${\mathcal{L}}_{max}$ respectively.
 For one parameter fitting, $-2\ln({\cal {L}}/{\mathcal{L}}_{max})$
 is distributed like a
 $\chi^{2}$ distribution with one degree of freedom. The
 $68\%$ and $95.4\%$ confidence limits on the parameter are where
 $\cal{L}$
 is $60.6\%$ and $14\%$ of ${\mathcal{L}}_{max}$ respectively
 (Kochanek
 (1993)\cite{1CSK} and  Lampton \emph{et al.} (1976)\cite{Xray})  .

 \vskip 0.3cm
 \noindent$\bullet$ {\bf No-Evolution Model\,:} Fig. 1 shows the
 contours of
 constant likelihood ($95.4\%$ and $68\%$) in the  two parameter space
 ($w$, $\Omega_{m}$). The best fit (${\mathcal{L}}_{max}$) occurs
 for $w = -0.33$
 and $\Omega_{m} = 0.0$. We see that $w \leq -0.04$ and
 $\Omega_{m}\leq 0.90$ at $1\sigma$ ($68\%$ confidence level).
 For the case of constant $\Lambda$ i.e. $w
 = -1.0$, ${\mathcal{L}}_{max}$ occurs for $\Omega_{m} = 0.44$
 ($\Omega_{\Lambda} = 0.56$)
 and $0.27\leq \Omega_{m}\leq 0.75$ at $1\sigma$. We get 
 $\Omega_{\Lambda} \leq 0.83$ at $2\sigma$ for $w = -1$. Waga \& Miceli
 \cite{Waga} have done a similar study with  Kochaneck's values
 \cite{2CSK} for
 Schechter and lens parameters. For a one parameter fit they get
 $\Omega_{\Lambda}
 \leq 0.55$ at $2 \sigma$. 
%These results clearly show that the limit
%on $w$ is sensitive to Lens and Schechter parameters. \\

 \noindent$\bullet$ {\bf VG Model\,:} Fig. 2 summarises the results
 for the VG
 Model. The maximum of likelihood occurs for $w =-0.33$ and
 $\Omega_{m} = 0.0$ with $w \leq -0.04$ and $\Omega_{m} \leq 0.91$
 at $1
 \sigma$. For the constant $\Lambda$ case, ${\mathcal{L}}_{max}$
 occurs for $\Omega_{m}
 = 0.44$. Also $0.26 \leq \Omega_{m} \leq 0.74 $ at $1\sigma$ and 
$\Omega_{m} \geq 0.17$ at $2\sigma$.\\

 \noindent$\bullet$ {\bf Fast Merging Model\,:} Fig. 3 shows the
 contours of constant
 likelihood ($68\%$ and $95.4\%$) for the Fast Merging Model of
 galaxy
 evolution. The best fit occurs for $w = -0.26$ and $\Omega_{m} =
 0.0$. For this model of galaxy evolution the constraints are
 very weak. For a one parameter fit i.e. for the case of constant
 $\Lambda$, ${\mathcal{L}}_{max}$
 occurs for $\Omega_{m} = 0.52$ and $0.29 \leq \Omega_{m} \leq 0.92$
 at $1 \sigma$ and $\Omega_{m} \geq 0.16$ at $2\sigma$.

 \vskip 0.2cm
 \noindent Table 1 summarises the results for the three models.

 \vskip 0.3cm

\noindent Torres and Waga \cite{Bloom}, Waga and Miceli \cite{Waga}
 have done a similar study for decaying $\Lambda$ cosmologies
 ($\rho_\Lambda \propto a^{-m}$) without taking evolution of galaxies into
 account.
 They point out the fact that the constraints obtained from
 gravitational lens statistics are very weak in these decaying $\Lambda$
cosmologies.
 The reason for this is  that the distance to an object
 is smaller (for fixed $z$ and $\Omega_{m}$ ) in  decaying $\Lambda$
 models as compared to the distance of the same object in constant
 $\Lambda$
 models for the same value of $z$ and $\Omega_{m}$. Hence the
 probability
 of lensing of that object is reduced in decaying $\Lambda $ model
 cosmologies. Therefore the constraints on $m$ from the statistics of lensing
 become weaker.

 In this paper our main aim is to check how the constraints on $w$
 changes when we include galaxy evolution into the lensing
 statistics ( in dynamical $\Lambda$
 models : X - matter cosmology i.e. $\rho_x
 \propto a^{-3(1 + w)}$) .
 It is a well known fact that the lensing probability depends upon the
 source redshift, luminosity of the quasar, lens model, the
 cosmological model and more importantly on the number density of
 galaxies (lenses). The evolution of galaxies changes the number
 density
 of galaxies and hence the probability of lensing changes.

 Fig. 4 illustrates how
 the  optical depth (lensing probability) decreases with increasing
 $w$. This is because as $w$ increases, the distance
 to an object at redshift $z$ decreases which in turn decreases the
 probability of lensing. The inclusion of galaxy evolution further
 decreases the lensing probability in dynamical $\Lambda$
 models ( X - matter cosmology ). As a result the constraints
 on the cosmological parameters ($w$, $\Omega_{m}$) become weaker.
 In the case of a two parameter fit, we observe that higher
 values of $w$ permit only low values of $\Omega_{m}$.

	The probability for a quasar to get lensed decreases marginally when we
go from no-evolution model to VG model for galaxy evolution
and hence the constraints on $w$ and $\Omega_{m}$ obtained 
in the case of the VG model are similar to those obtained 
in the no-evolution model. On the other hand, the probability 
for a quasar to get lensed
decreases considerably when we work with Fast Merging model for galaxy evolution
This gets reflected as weaker constraints in the Fast Merging model. 
Jain \emph{et al.} \cite{DJ1}
constrained $\Omega_{\Lambda}$ (constant $\Lambda$) for flat
 cosmologies taking into account galaxy evolution. They worked with
 the
 BFSP quasar sample. They constrained it
 using the information of the number of lensed quasars and concluded
 that
 evolution permits a larger value of $\Omega_{\Lambda}$. They also
 found that in their formalism $\gamma = 4$ masks the difference
 which
 the  evolution of galaxies makes. We would like to point out
 that the present formalism allows us to study the difference which
 comes in when the evolution of galaxies is taken into account even with
$\gamma = 4$.

 It is interesting to note that the image separation distribution
 is different for the three models (Fig. 5). This image separation
 is
 sensitive to lens, Schechter and cosmological parameters
 \cite{dj}. Fig. 6 gives the predicted number of lenses
 $N_{lens}$ in the adopted quasar sample for different $\Omega_{m}$
 values ($w = -1$ case) in the three models.
 The results are similar to those obtained by Jain \emph{et al.}.
 i.e. \cite{DJ1}the evolutionary models permit larger values of
 $\Omega_{\Lambda}$. The constraint that in the given optical sample
 $N_{L} = 5$ gives us different values of $\Omega_{\Lambda}$ in different
 models.

 \begin{enumerate}
 \item No evolution: $\Omega_{\Lambda} \sim 0.54$
 \item VG model: $\Omega_{\Lambda} \sim 0.56$
 \item Fast merging model: $\Omega_{\Lambda} \sim 0.60$
 \end{enumerate}

 In the calculation of number of lensed quasars, $N_L$ in the
 given sample (Fig.6), no information of image morphology is 
 taken into account as explained in
 Sec. 4.1.  Likelihood analysis, on the other hand, takes into account
 the observed image separations.
 There is not much difference in the predicted values of 
 $\Omega_{\Lambda}$ by these two methods in the case of no
 evolution and the VG model for  $( w = -1)$ as can be seen 
 by comparing the above mentioned values to those in  Table 1.
 However, for the constant $\Lambda$ case, likelihood
 analysis shows that
 the Fast Merging model permits higher values of $\Omega_{m}$ .
 Calculating the predicted number of  lensed quasars and constraining the
 parameters using the observational fact that in the adopted quasar
 sample there are five lenses  does not make use of the information
 about the image separations observed for the lensed quasars.
 The results of likelihood analysis, however,  crucially depend
 on the information about image separation. We feel that it is
 important to incorporate the  image
 separations information while predicting (constraining) the cosmological
 parameters. 

The constraints obtained from lensing statistics are sensitive to 
both lens and Schechter parameters \cite{c,DA}.
For the quasar luminosity function
used in this paper, it is observed that for the case 
of one parameter fit ( $w = -1$) K96 parameters (Table 2) give  $N_{L} = 5$ 
( for this optical sample of 867 quasars) only when $\Omega_{m}$ is nearly 
equal to one. For all other values of $\Omega_{m}$, K96 parameters give
$N_{L}$ greater than the observed number of lensed quasars (there
are five observed lensed quasars) in this optical quasar sample.
The LPEM parameters, on the other hand, give numbers around
this observed number of lensed quasars over a wide range of
parameter ($\Omega_{m}$) values.
We have therefore restricted ourselves to the use of 
LPEM lens and Schechter parameters to study how the galaxy evolution models
change the constraints on the cosmological paramters.
We observe that the probability for a quasar to get lensed 
increases with decrease in $\Omega_{m}$ or/and $w$. 
Therefore for small values of $\Omega_{m}$ and $w$ we cannot use 
the approximation $\ln(1-p_{i}^{'})\sim -p_{i}^{'}$ (in eq. \ref{LLF})
which holds when $p_{i}^{'} \ll 1$. Infact, for consistency, 
we have not worked with this approximation for the entire range 
of the cosmological parameters.

It is interesting to note that with K96 parameters, the lensing
probability of a few bright quasars at high redshift becomes greater than one for
$\Omega_{m} \sim 0.0$ and $w \sim -1$. 
For instance, one quasar in the sample with z = 3.02 and $m_{B} = 16.1$ 
gives a probablity($p_i^{'}$) of 1.42 for $\Omega_{m} = 0.0$ and $w = -1$. 
This is not an artifact of our numerical
programming and can be checked easily by using either eq. (\ref{prob2}) of this paper
or eq. (4.15) of K93 \cite{1CSK}.  
It may be because of the specific values of parameters 
that we have taken. Specifically, the value of $F^{*}$  for K96 parameters 
($F^{*}=0.026$) is quite large as   
compared to that for the LPEM parameters ($F^{*}=0.010$). 
Clearly this larger value of $F^{*}$ pushes up the 
probabilities  significantly. In fact, even for those sample 
points where the probability is less than one, there are instances 
where the probability is large enough ( near $\Omega_m \sim 0.0$ ) so that the 
approximation  $\ln(1-p_{i}^{'})\sim -p_{i}^{'}$  is invalid.

It could also be that the  formalism
itself is inconsistent, though there is no obvious 
inconsistency when parameters other than K96 are used. 
If the problem is that of the formalism not being consistent, 
then one would need to investigate 
exactly what is it in the formalism which leads to these 
seemingly absurd probability values. 

The strength of the
constraints obtained also depends strongly on the lensing data. 
Extended surveys are required to
 establish $n(\Delta\theta)$ as a powerful tool for constraining 
the parameters. The upcoming Sloan Digital
 Sky
 Survey which is going down to $1 - \sigma$ magnitude limit of $\sim
 23$ will definitely increase our understanding of lensing
 phenomena and  cosmological parameters. A large number of
 new gravitational lens systems $(\sim 18)$ have been discovered by the
 Cosmic Lens All-Sky Survey (CLASS)\cite{tch,rt} . CLASS is the
 largest survey  
 with more than 10,000 radio sources down to a 5 GHz flux
 density of 30 mJy. But the major disadvantage of a radio lens survey is
 that there is little information on the redshift-dependent
 number-magnitude relation. This leads to serious
 systematic uncertainties in the derived cosmological constraints.

 In this work we have included the effects  of galaxy evolution to  
  constrain  the cosmological parameters
 ($w$, $\Omega_{m}$) obtained from lensing statistics. We find
 that while the constraints obtained in the case of VG model are
 similar to
 those obtained for the no evolution case, they  are much
 weaker
 in the case of the Fast Merging model.
 We have only considered
 those galaxy evolutionary models which are based on the assumption
 that the total comoving mass is conserved during mergers. There is
 a possibility that the total comoving mass isn't  conserved when
 mergers take place. Work is under progress to see how lensing
 statistics and cosmological parameters are affected when total
 comoving mass density of lenses is not conserved.

\end{section}

\begin{section}*{Acknowledgements}
   %~~~~~~~~~~~~~~~~~~~~~~~~~~~~~~~~~~~~~~~~~~~~~~~~
%              ACKNOWLEDGEMENT
%~~~~~~~~~~~~~~~~~~~~~~~~~~~~~~~~~~~~~~~~~~~~~~~~

We are thankful to I. Waga for providing us the quasar sample.
The authors are  greatful to C. Kochanek, I. Waga and A. Mukherjee 
for useful discussions during the course of this work.
One of the author (A. Dev) thanks University Grant Commission of India for 
providing research fellowship.

   \end {section}

\vfill
\eject

\begin{table}
\begin{center}
\begin{tabular}{|l|l|l|l|r|} \hline
Model & Fit & Best Fit & $68\%$ CL & $95.4\%$ CL \\ 
&   & $w$, $\Omega_{m}$ & & \\ \hline
No-Evolution & A & -0.33, 0.0& $w \leq -0.04$  &   \\
                 &  &          & $\Omega_{m} \leq 0.90$ &  \\
                 & B & 0.44 & $0.27\leq\Omega_{m}\leq 0.75$ &
$0.17\leq\Omega_{m}$ \\ \hline
VG               & A & -0.33, 0.0& $w \leq -0.04$  &  \\
                 &  &          & $\Omega_{m} \leq 0.91$ & \\
                 & B & 0.44 & $0.26\leq\Omega_{m}\leq 0.74$ &
$0.17\leq\Omega_{m}$ \\ \hline
Fast Merging & A & -0.26, 0.0&    &  \\
                 &  &          &  & \\
                 & B & 0.52 & $0.29\leq\Omega_{m}\leq 0.92$ &
$0.16\leq \Omega_{m}$ \\ \hline

\end{tabular}

\caption{ Here  A refers to the two parameter fit and B refers to
one parameter fit i.e. when $w =-1$.}
\end{center}
\end{table}

\begin{table}
\begin{center}
%  TABLE 2 : LENS AND SHECTER PARAMETERS

\begin{tabular}{|l|l|l|l|l|r|} \hline
Survey & $\alpha$ & $\phi_{\ast}(h^{3}\,Mpc^{-3})$  & $v_{\ast}(km/s)$ & $F^{*}$\\ \hline
LPEM   &  +0.2   & $3.2 \times 10^{-3}$  & 205.3  & 0.010      \\ 
       &         &                       &        &             \\ \hline
K96    &  -1.0  & $6.1 \times 10^{-3} $ & 225.0  & 0.026       \\ 
       &         &                       &        &             \\ \hline
\end{tabular}

\caption{ Lens and Schecter Paramters for E/SO galaxies }
\end{center}
\end{table}

\begin{figure}[ht]
\vskip 15 truecm
\includegraphics{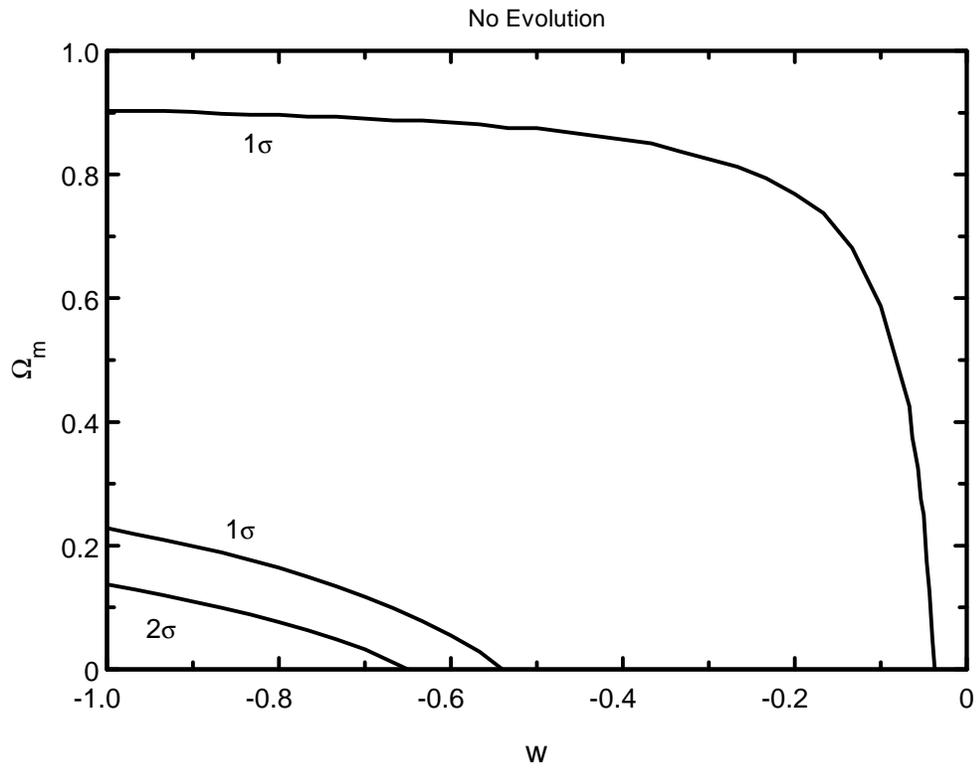}
\caption{ Contours of constant likelihood ($68\%$ and $95.4\%$) for 
No-Evolution Model for lensing galaxies}
\end{figure}

\begin{figure}[ht]
\vskip 15 truecm
\includegraphics{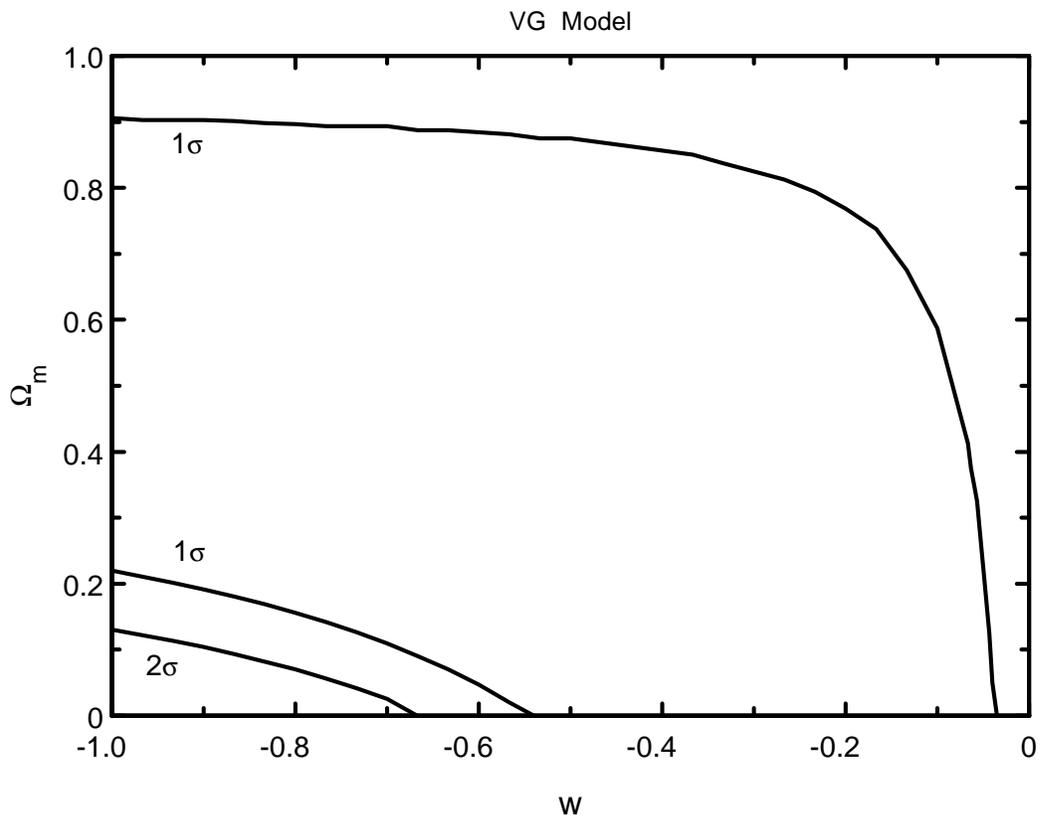}
\caption{ Contours of constant likelihood ($68\%$ and $95.4\%$) for
Volmerange \& Guiderdoni Model of galaxy evolution.}
\end{figure}
\vfill
\eject

\begin{figure}[ht]
\vskip 15 truecm
\includegraphics{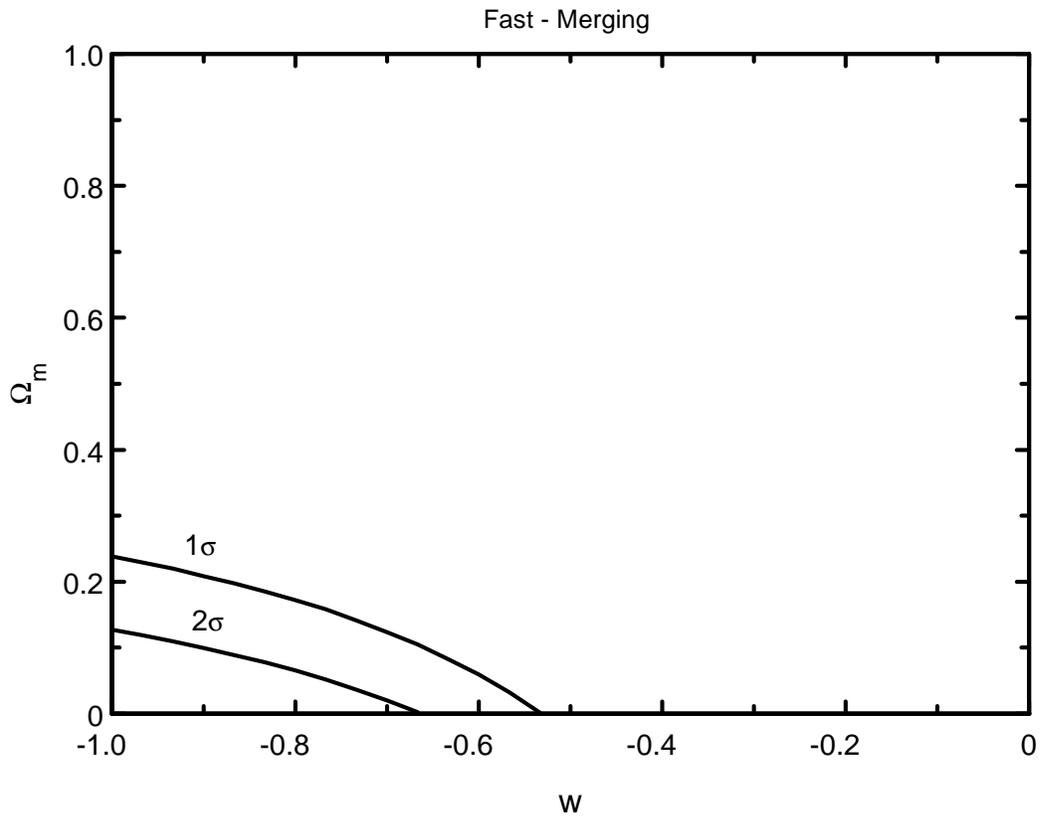}
\caption{Contours of constant likelihood ($68\%$ and $95.4\%$) for
Fast Merging Model of galaxy evolution.}
\end{figure}
\vfill
\eject

\begin{figure}[ht]
\vskip 15 truecm
\includegraphics{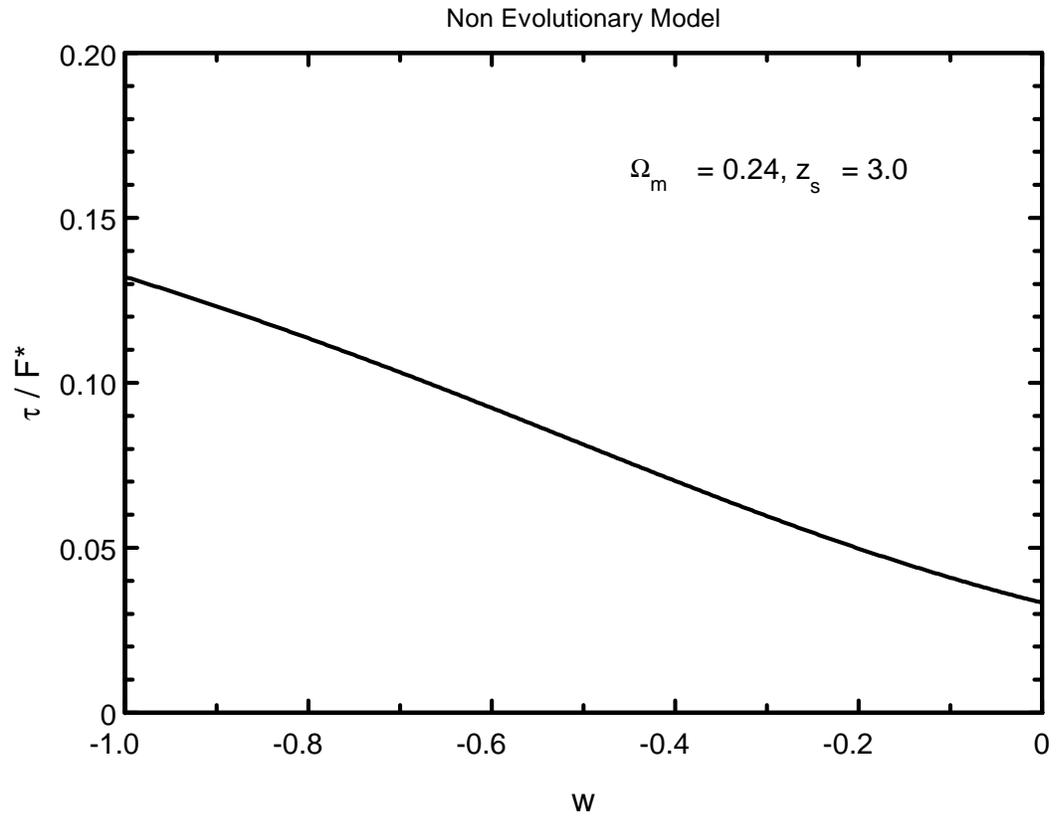}
\caption{Probability of lensing as a function of $w$}
\end{figure}
\vfill
\eject

\begin{figure}[ht]
\vskip 15 truecm
\includegraphics{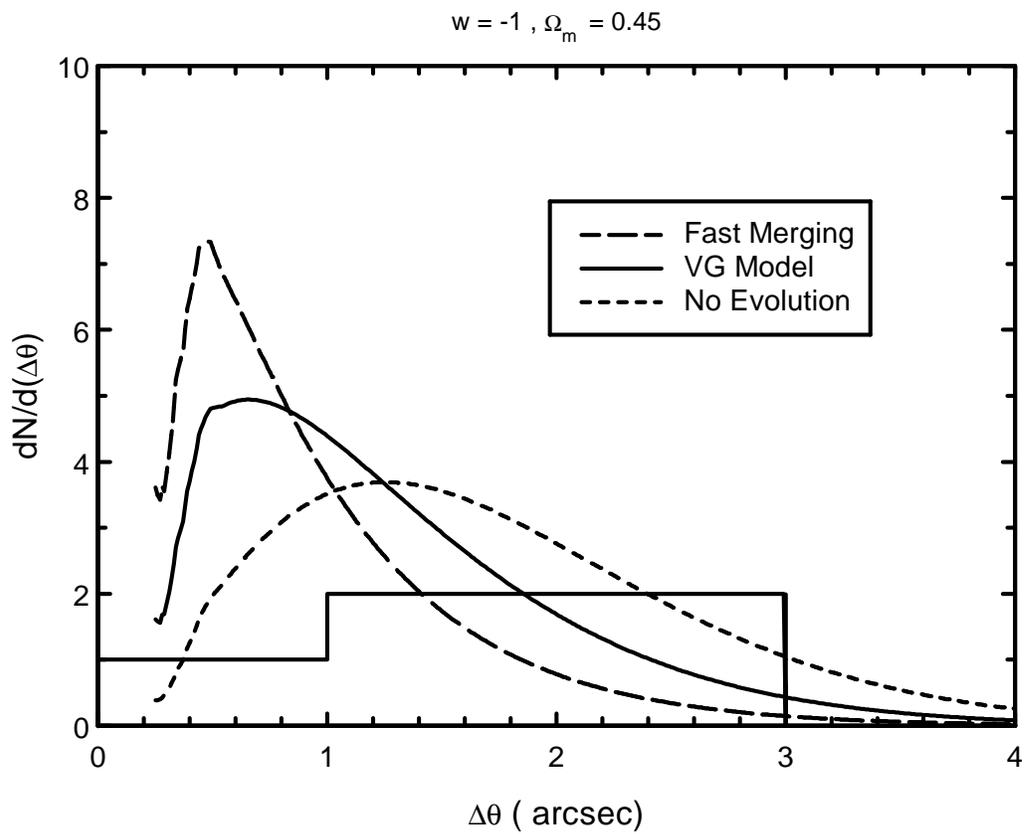}
\caption{Predicted image-separation distribution $dN/d(\Delta\theta)$ compared with
the observed image-separation in the optical quasar sample(histogram).}
\end{figure}
\vfill
\eject

\begin{figure}[ht]
\vskip 15 truecm
\includegraphics{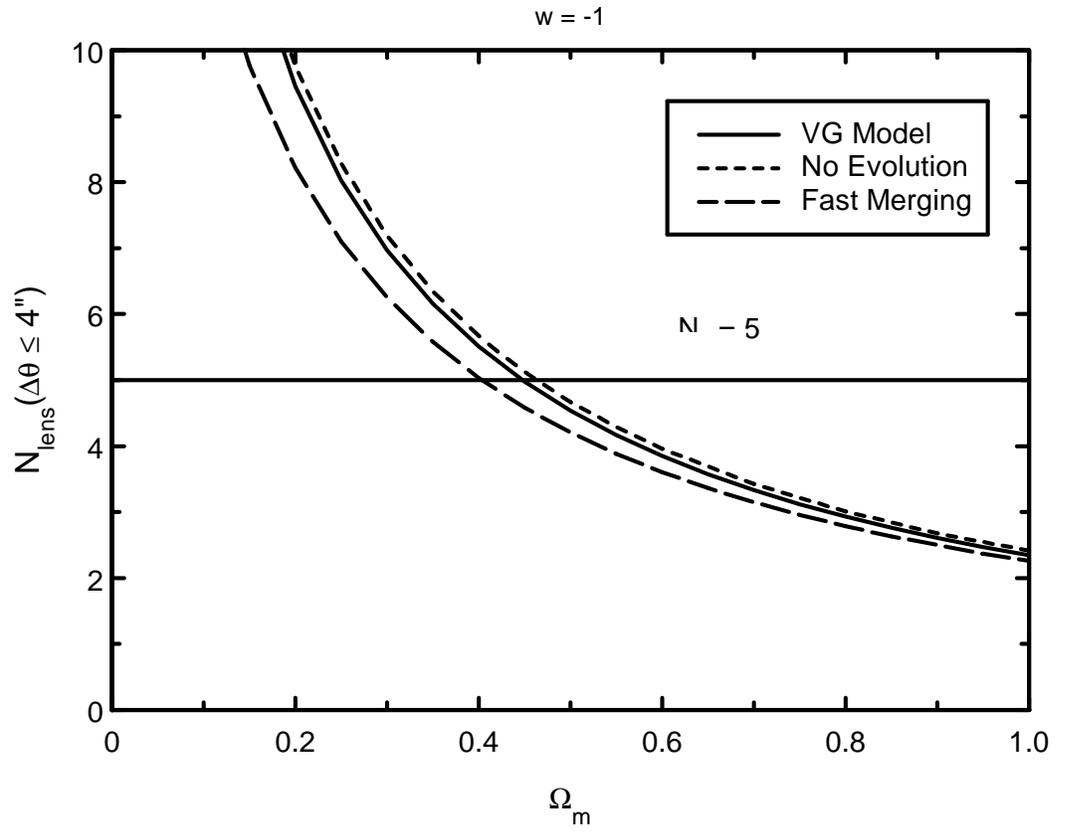}
\caption{Pridicted number of lensed quasars with $\Delta\theta \leq 4''$  
 in the adopted optical quasars sample.}
\end{figure}
\vfill
\eject

\end{document}